\documentclass[%
 reprint,
nofootinbib,
 amsmath,amssymb,
 aps,
]{revtex4-2}

\usepackage{xcolor}% Color text
\usepackage{multirow}
\usepackage{graphicx}% Include figure files
\usepackage{dcolumn}% Align table columns on decimal point
\usepackage{bm}% bold math
\usepackage{hyperref}% add hypertext capabilities
\newcommand{\wmap}[0]{\textit{WMAP}}
\newcommand{\planck}[0]{\textit{Planck}}

\newcommand{\npipe}[0]{\texttt{NPIPE}}

\begin{document}

%\preprint{APS/123-QED}

\title{Improved Constraints on Cosmic Birefringence from the \wmap\ and \planck\ Cosmic Microwave Background Polarization Data}

\author{Johannes R. Eskilt}
\email{j.r.eskilt@astro.uio.no}
\affiliation{Institute of Theoretical Astrophysics, University of Oslo, P.O. Box 1029 Blindern, N-0315 Oslo, Norway}
\author{Eiichiro Komatsu}
\email{komatsu@mpa-garching.mpg.de}
\affiliation{
Max Planck Institute for Astrophysics, Karl-Schwarzschild-Str. 1, D-85748 Garching, Germany
}
\affiliation{Kavli Institute for the Physics and Mathematics of the Universe (Kavli IPMU, WPI), Todai Institutes for Advanced Study, The University of Tokyo, Kashiwa 277-8583, Japan}

\date{\today}% It is always \today, today,
\begin{abstract}
The observed pattern of linear polarization of the cosmic microwave background (CMB) photons is a sensitive probe of physics violating parity symmetry under inversion of spatial coordinates. A new parity-violating interaction might have rotated the plane of linear polarization by an angle $\beta$ as the CMB photons have been traveling for more than 13 billion years. This effect is known as ``cosmic birefringence.'' In this paper, we present new measurements of cosmic birefringence from a joint analysis of polarization data from two space missions, \planck\ and \wmap. This dataset covers a wide range of frequencies from 23 to 353~GHz. We measure $\beta = 0.342^{\circ\,+0.094^\circ}_{\phantom{\circ\,}-0.091^\circ}$ (68\% \ \textrm{C.L.}) for nearly full-sky data, which excludes $\beta=0$ at 99.987\%~C.L. This corresponds to the statistical significance of $3.6\sigma$. There is no evidence for frequency dependence of $\beta$. We find a similar result, albeit with a larger uncertainty, when removing the Galactic plane from the analysis.
\end{abstract}

\maketitle

\section{\label{sec:intro}Introduction}

Photons of the cosmic microwave background (CMB), the afterglow of the primordial fireball Universe~\cite{peebles/page/partridge:2009}, are linearly polarized~\cite{kosowsky:1996}. One can decompose the observed pattern of linear polarization into eigenstates of parity called $E$ and $B$ modes, which transform differently under inversion of spatial coordinates~\cite{zaldarriaga/seljak:1997,kamionkowski/etal:1997}. This property can be used to probe new physics beyond the standard model of elementary particles and fields that violates parity symmetry~\cite{lue/wang/kamionkowski:1999}.

A pseudoscalar ``axionlike'' field, $\phi$, that couples to electromagnetism is an example of such new physics. Consider a Lagrangian density given by~\cite{ni:1977,turner/widrow:1988}
\begin{equation}
\label{eq:lagrangian}
    {\mathcal L}=-\frac12(\partial\phi)^2-V(\phi)-\frac14F^2-\frac{\alpha}{4f}\phi F\tilde F\,,
\end{equation}
where $(\partial\phi)^2\equiv g^{\mu\nu}\partial_\mu\phi\partial_\nu\phi$ with the metric tensor $g^{\mu\nu}$, $V(\phi)$ is $\phi$'s potential,
$F^2\equiv g^{\mu\mu'}g^{\nu\nu'}F_{\mu\nu}F_{\mu'\nu'}$, $F\tilde F\equiv F_{\mu\nu}\epsilon^{\mu\nu\mu'\nu'}F_{\mu'\nu'}/(2\sqrt{-g})$, and $\alpha$ and $f$ are a dimensionless coupling constant and the so-called axion decay constant, respectively. Here, $F_{\mu\nu}\equiv\partial_\mu A_\nu-\partial_\nu A_\mu$ is the antisymmetric electromagnetic tensor with the vector potential $A_\mu$, $\epsilon^{\mu\nu\mu'\nu'}$ is a totally antisymmetric symbol with $\epsilon^{0123}=1$, and $g$ is the determinant of the metric tensor. 

The last term in Eq.~\eqref{eq:lagrangian} is familiar in particle physics, as it appears in the low-energy effective action of quantum chromodynamics~\cite{weinberg:1996}. A pion is a pseudoscalar~\cite{chinowsky/steinberger:1954} and a neutral pion decays into two photons via this term with $\alpha$ and $f$ determined precisely by experiments. In cosmology, $\phi$ is a new pseudoscalar and a candidate for dark matter and dark energy~\cite{marsh:2016,ferreira:2021} with $\alpha/f$ being a free parameter. 

When $\phi$ depends on spacetime, the plane of linear polarization of photons rotates~\cite{carroll/field/jackiw:1990,carroll/field:1991,harari/sikivie:1992} by an angle
$\beta(\hat{\mathbf n})
    =\frac{\alpha}{2f}\left[\phi(\eta_{\textrm{o}})-\phi(\eta_{\textrm{e}},r\hat{\mathbf n})\right]$, where $\hat{\mathbf n}$ is the direction of an observer's line of sight, $\eta_{\textrm{o}}$ and $\eta_{\textrm{e}}$ are the conformal times of observation and emission of photons, respectively, and $r\equiv\eta_{\textrm{o}}-\eta_{\textrm{e}}$ is the conformal distance to the emitter. We take the speed of light to be unity throughout this paper.
    
This effect is often called ``cosmic birefringence,'' and is best probed by analyzing the oldest polarized light in the Universe, that is, the CMB (see Ref.~\cite{komatsu:2022} for a review). Discovery of a non-zero value of $\beta$ would have profound implications for the fundamental physics behind dark energy~\cite{carroll:1998,panda/sumitomo/trivedi:2011}, dark matter~\cite{finelli/galaverni:2009,fedderke/graham/rajendran:2019}, and quantum gravity~\cite{myers/pospelov:2003,arvanitaki/etal:2010}. 

When analyzing CMB data, one decomposes the Stokes parameters for linear polarization as~\cite{zaldarriaga/seljak:1997,kamionkowski/etal:1997}
\begin{equation}
    Q(\hat{\mathbf n})\pm iU(\hat{\mathbf n})=-\sum_{\ell=2}^{\ell_\mathrm{max}}\sum_{m=-\ell}^\ell\left(E_{\ell m}\pm iB_{\ell m}\right){}_{\pm 2}Y_\ell^m(\hat{\mathbf n})\,,
\end{equation}
where $E_{\ell m}$ and $B_{\ell m}$ are the spherical harmonics coefficients of the $E$ and $B$ modes, respectively, ${}_{\pm 2}Y_\ell^m(\hat{\mathbf n})$ are the spin-2 spherical harmonics, and $\ell_\mathrm{max}$ is the maximum multipole used for the analysis. 

The coefficients transform under inversion of spatial coordinates, $\hat{\mathbf n}\to -\hat{\mathbf n}$, as $E_{\ell m}\to (-1)^\ell E_{\ell m}$ and $B_{\ell m}\to (-1)^{\ell+1} B_{\ell m}$. The cross-power spectrum of $E$ and $B$ modes, $C_\ell^{EB}\equiv (2\ell+1)^{-1}\sum_{m}\operatorname{Re}(E_{\ell m}B_{\ell m}^*)$, has odd parity and is sensitive to $\beta$~\cite{lue/wang/kamionkowski:1999}.
When the plane of linear polarization rotates uniformly on the sky by an angle $\beta$, the observed $E$ and $B$ modes become
$E_{\ell m}^{\textrm{o}}=E_{\ell m}\cos(2\beta)-B_{\ell m}\sin(2\beta)$ and $B_{\ell m}^{\textrm{o}}=E_{\ell m}\sin(2\beta)+B_{\ell m}\cos(2\beta)$, respectively. One thus finds that~\cite{liu/lee/ng:2006,feng/etal:2005,zhao/etal:2015,gruppuso/etal:2016,minami/etal:2019}
\begin{equation}
    \label{eq:simple_eb}
    C_\ell^{EB, {\textrm{o}}} = \frac{\tan(4\beta)}{2}\left( C^{EE, {\textrm{o}}}_\ell - C^{BB, {\textrm{o}}}_\ell\right)+\frac{C_\ell^{EB}}{\cos(4\beta)}\,,
\end{equation}
where the last term is the intrinsic $EB$ correlation at the time of emission, and $C_\ell^{EE}$ and $C_\ell^{BB}$ are the auto-power spectra of $E$ and $B$ modes, respectively. We adopt the sign convention such that $\beta>0$ is a clockwise rotation of the plane of linear polarization on the sky.

However, miscalibration angles will arise if one does not know precisely enough how the polarization-sensitive
orientations of detectors on the focal plane of a telescope are related to the sky coordinates and how polarization of the incoming light is rotated by optical components~\cite{wu/etal:2009,miller:2009,WMAP:2011,RAC:2022}. These angles will be degenerate with the cosmic birefringence angle $\beta$. We will from now on denote miscalibration angles by $\alpha$, which should not be confused with the coupling constant in Eq.~\eqref{eq:lagrangian}. Hence, the observed $EB$ power spectrum measured by an instrument gets an extra rotation contribution from its miscalibration angle $\alpha$, inducing a total rotation of $\alpha + \beta$. Without knowledge of $\alpha$, one can only determine the sum of the two angles, $\alpha+\beta$.

The sky contains not only the CMB, but also the polarized emission of interstellar gas, called the Galactic foreground emission.
Photons of the foreground emission do not travel for a long distance, receiving only a negligible amount of $\beta$ when $\phi$ varies slowly in spacetime; thus, the foreground polarization is rotated only by the miscalibration angle $\alpha$~\cite{minami/etal:2019}. The foreground emission might possess a non-vanishing intrinsic $EB$ correlation, which needs to be taken into account.

These considerations lead to
\begin{align}
    \nonumber
    \begin{bmatrix}
    E^{\textrm{o}}_{\ell m} \\
    B^{\textrm{o}}_{\ell m}
    \end{bmatrix} &= \begin{bmatrix}
    \cos(2\alpha) & -\sin(2\alpha)\\
    \sin(2\alpha) & \phantom{-}\cos(2\alpha)
    \end{bmatrix}\begin{bmatrix}
    E^{\textrm{fg}}_{\ell m} \\
    B^{\textrm{fg}}_{\ell m}
    \end{bmatrix}\\ 
    \label{eq:e_b_alm}
    &+ \begin{bmatrix}
    \cos(2\alpha+2\beta) & -\sin(2\alpha+2\beta)\\
    \sin(2\alpha+2\beta) & \phantom{-}\cos(2\alpha+2\beta)
    \end{bmatrix}\begin{bmatrix}
    E^{\textrm{CMB}}_{\ell m} \\
    B^{\textrm{CMB}}_{\ell m}
    \end{bmatrix}\,,
\end{align}
for a single channel. Here, ``fg'' and ``CMB'' denote the foreground and CMB, respectively.
One finds that \cite{minami/etal:2019}
\begin{align}
    \nonumber
    C_\ell^{EB, {\textrm{o}}} &= \frac{\tan(4\alpha)}{2}\left(C_\ell^{EE, {\textrm{o}}}- C_\ell^{BB, {\textrm{o}}}\right)+ \frac{C_\ell^{EB, \text{fg}}}{\cos(4\alpha)} \\
    \nonumber
    &+ \frac{\sin(4\beta)}{2\cos(4\alpha)}\left(C_\ell^{EE, \text{CMB}}- C_\ell^{BB, \text{CMB}} \right)\\
    \label{eq:singleinstrument}
    &+ \frac{\cos(4\beta)}{\cos(4\alpha)} C_\ell^{EB, \text{CMB}}\,.
\end{align}
This equation allows us to determine $\alpha$ and $\beta$ simultaneously, independent of the $E$- and $B$-mode auto-power spectra of the  foreground, $C^{EE, \textrm{fg}}_\ell$ and $C^{BB, \textrm{fg}}_\ell$. 

Eq.~\eqref{eq:singleinstrument} still requires knowledge of 
$C_\ell^{EB,\mathrm{fg}}$ and $C_\ell^{EB,\mathrm{CMB}}$. Discovery of the latter would be similarly revolutionary in cosmology; however, we ignore this term in this paper because the current data are not yet sensitive enough to detect it. 

We take into account the effect of $C_\ell^{EB,\mathrm{fg}}$. As the foreground helps us constrain $\alpha$, neglecting any non-zero $C_\ell^{EB,\mathrm{fg}}$ in the analysis can bias the measurement of $\alpha$. A biased $\alpha$ will necessarily bias $\beta$ since the CMB highly constrains the sum, $\alpha+\beta$.

After being verified on simulations with multiple frequency channels \cite{minami/komatsu:2020}, the method was applied to the high frequency instrument (HFI~\cite{Planck2018III}) maps of the \planck\ Public Release 3 (PR3) in Ref.~\cite{minami/komatsu:2020b}. The authors measured $\beta = 0.35^\circ \pm 0.14^\circ$ for nearly full-sky data. We quote the 68\% confidence level (C.L.) intervals throughout this paper. The method was then applied to the \planck\ PR4 HFI data~\cite{PlanckIntLVII}, yielding $\beta=0.30^\circ \pm 0.11^\circ$~\cite{NPIPE:2022}. The low frequency instrument (LFI~\cite{Planck2018II}) maps of the \planck\ PR4 were included in the analysis of Ref.~\cite{Eskilt:2022wav}, which found $\beta=0.33^\circ \pm 0.10^\circ$. The statistical significance exceeds $3\sigma$.

Ref.~\cite{Eskilt:2022wav} also measured the frequency dependence of the signal by fitting a power-law model, $\beta(\nu) \propto \nu^n$, finding $n=-0.35^{+0.48}_{-0.47}$. This is consistent with a frequency-independent signal predicted by the axionlike field, Eq.~\eqref{eq:lagrangian}.

These authors \cite{minami/komatsu:2020b,NPIPE:2022,Eskilt:2022wav} initially ignored $C_\ell^{EB,\mathrm{fg}}$ in the analysis. As the foreground is dominated by polarized thermal dust emission at the HFI frequencies~\cite{Planck2018XI}, they noted that the positive dust $TB$ and $TE$ correlations found from \planck\ \cite{Planck2018XI} would suggest $C_\ell^{EB,\mathrm{fg}}\simeq C_\ell^{EB,\mathrm{dust}}>0$, which would make the measured value of $\beta$ a lower bound. This expectation was confirmed by Ref.~\cite{NPIPE:2022}, which found a larger value, $\beta=0.36^\circ \pm 0.11^\circ$, from nearly full-sky data of the \planck\ HFI PR4 assuming $C_\ell^{EB,\mathrm{dust}}/C_\ell^{EE,\mathrm{dust}}\propto C_\ell^{TB,\mathrm{dust}}/C_\ell^{TE,\mathrm{dust}}$~\cite{clark/etal:2021}.

When $C_\ell^{EB,\mathrm{dust}}$ is ignored, the inferred value of $\beta$ decreases as the Galactic plane is masked and removed from the analysis via the impact of $C_\ell^{EB,\mathrm{dust}}$ on the determination of $\alpha$~\cite{NPIPE:2022}. Although one finds similar values of $\beta$ from the \planck\ data regardless of the sky fraction used for the analysis when $\alpha=0$, that is, the $EB$ signal of $\alpha+\beta\simeq 0.3^\circ$ is isotropic in the sky, the inferred value of $\alpha$ depends on the sky fraction via $C_\ell^{EB,\mathrm{dust}}$. This was foreseen in the work by Ref.~\cite{clark/etal:2021}, which argued that there should be a mostly positive  $C_\ell^{EB,\mathrm{dust}}$ for smaller sky fractions (larger Galactic masks). This would bias a measurement of $\beta$ towards a lower value whenever $C_\ell^{EB,\mathrm{dust}}$ is ignored.

The authors of Ref.~\cite{NPIPE:2022} suggested an ansatz to model $C_\ell^{EB,\mathrm{dust}}$ based on the results of Ref.~\cite{huffenberger/rotti/collins:2020,clark/etal:2021}. Including their ansatz to the equations confirmed that $C_\ell^{EB,\mathrm{dust}}$ was the cause of the decline of $\beta$ for lower sky fractions. Including $C_\ell^{EB,\mathrm{dust}}$ in the inference gave robust positive measurements of $\beta$ at all sky fractions.

Therefore, the results so far indicate the presence of an isotropic and frequency-independent signal of cosmic birefringence in the \planck\ data with a statistical significance of $3\sigma$. The detailed study~\cite{NPIPE:2022,diego-palazuelos/etal:2022} using the simulations of PR4~\cite{PlanckIntLVII} shows that the impact of the known systematics of the \planck\ HFI on $\beta$ is negligible compared with the statistical uncertainty.

In this paper, we continue to search for the isotropic $\beta$ by including the polarization data of the \textit{Wilkinson Microwave Anisotropy Probe} (\wmap) 9-year observations~\cite{WMAP:2013a} in a joint analysis with the polarized \planck\ channels. The polarization data of \wmap\  have a lower signal-to-noise ratio than those of the \planck\ HFI, but the inclusion of the \wmap\ channels gives rise to many cross-power spectra with both itself and the \planck\ LFI and HFI channels. This allows us to increase the precision of $\beta$.
As there is no evidence~\cite{contreras/boubel/scott:2017,bianchini/etal:2020,namikawa/etal:2020,gruppuso/etal:2020} for fluctuations in $\beta(\hat{\mathbf n})$, we focus on the isotropic $\beta$.

The rest of the paper is organized as follows. We describe the data and the analysis method in Sec.~\ref{sec:pipeline}. We present the  results in Sec.~\ref{sec:results} and conclude in Sec.~\ref{sec:conclusion}.

\section{\label{sec:pipeline}Data and Analysis method}

We use the \wmap\ 9-year maps for each differencing assembly, which contain 1, 1, 2, 2, and 4 maps at $\nu=23$, 33, 41, 61, and 94~GHz, respectively~\cite{WMAP:2013a}. As in Refs.~\cite{NPIPE:2022,Eskilt:2022wav} we also use the \planck\ PR4 (often called \npipe-reprocessed data~\cite{PlanckIntLVII}) maps at $\nu=30$, 44, 70, 100, 143, 217, and 353~GHz. The \npipe\ pipeline divided detectors of a given frequency band into two groups, hence creating 2 detector split maps for each band. However, as there were not enough detectors to make 2 maps for each of 30 and 44\, GHz bands, we use the so-called half-mission maps for these bands. This means that the 30 and 44\,GHz bands have 1 miscalibration angle each, whereas the others have 1 miscalibration angle per detector split map.

The analysis method we use in this paper is similar to those presented in Refs.~\cite{minami/komatsu:2020b,NPIPE:2022,Eskilt:2022wav}, which we summarize here. The multi-channel generalization of Eq.~\eqref{eq:singleinstrument} is
\begin{align}
\nonumber
& C_\ell^{E_iB_j, \textrm{o}} = R^T(\alpha_i, \alpha_j)\textbf{R}^{-1}(\alpha_i, \alpha_j) \begin{bmatrix}
       C_\ell^{E_iE_j, \textrm{o}}\\
       C_\ell^{B_iB_j, \textrm{o}}
    \end{bmatrix}+\\
    \nonumber
    &\bigg[R^T(\alpha_i+\beta_i,\alpha_j+\beta_j)-R^T(\alpha_i,\alpha_j)\textbf{R}^{-1}(\alpha_i, \alpha_j)\\
    \label{eq:full_equation}
    &\cdot\textbf{R}(\alpha_i+\beta_i, \alpha_j+\beta_j) \bigg]\begin{bmatrix}
        C_\ell^{E_iE_j, \text{CMB}}\\
        C_\ell^{B_iB_j, \text{CMB}}
        \end{bmatrix}\,,
\end{align}
where
\begin{align}
    \mathbf{R}(\theta_i, \theta_j) &= \begin{bmatrix}
        \cos(2\theta_i)\cos(2\theta_j) & \sin(2\theta_i)\sin(2\theta_j) \\
        \sin(2\theta_i)\sin(2\theta_j) & \cos(2\theta_i)\cos(2\theta_j)
    \end{bmatrix}\,,\\
    R(\theta_i, \theta_j) &= \begin{bmatrix}
        \phantom{-}\cos(2\theta_i)\sin(2\theta_j)\\
        -\sin(2\theta_i)\cos(2\theta_j)
    \end{bmatrix}\,.
\end{align}
Here, $\alpha_i$ is the miscalibration angle for a given frequency band and data split, $i$. As we also allow $\beta$ to depend on frequency, $\beta_i$ denotes the value of $\beta(\nu)$ at $\nu=\nu_i$. We group the cross-power spectra of different combinations of maps into the observed power spectra vector $\vec{C}^{\textrm{o}}_\ell = \begin{bmatrix}
  C^{E_i E_j, \textrm{o}}_\ell, C^{B_i B_j, \textrm{o}}_\ell, C^{E_i B_j, \textrm{o}}_\ell
\end{bmatrix}^T$. The CMB power spectra, $C_\ell^{E_iE_j, \text{CMB}}$ and $C_\ell^{B_iB_j, \text{CMB}}$, are computed from the Boltzmann solver \texttt{CAMB}\cite{Lewis:2000}\footnote{\url{https://github.com/cmbant/CAMB}} with the best-fitting cosmological parameters given in Ref.~\citep{Planck2018VI}. We beam-smooth them using the beam transfer functions, $b^{X}_\ell$, and pixel window functions, $w^i_{\textrm{pix}}$,
\begin{equation}
    \vec{C}^{\Lambda \text{CDM}}_\ell = \begin{bmatrix}
    C^{EE,\text{CAMB}}_{\ell}  b^{E_i}_\ell b^{E_j}_\ell w^i_{\text{pix}, \ell}w^j_{\text{pix}, \ell}, \\C^{BB,\text{CAMB}}_{\ell} b^{B_i}_\ell b^{B_j}_\ell w^i_{\text{pix}, \ell}w^j_{\text{pix}, \ell}
    \end{bmatrix},
\end{equation}
where $C^{EE,\text{CAMB}}_{\ell}$ and $C^{BB,\text{CAMB}}_{\ell}$ are the power spectra from \texttt{CAMB}. As not all the official transfer functions of \wmap\ reach up to $\ell_{\textrm{max}}=1490$ used in our analysis, we set $b^{X}_\ell = 0$ for $\ell$ in which transfer functions are not available.

Following Refs.~\cite{huffenberger/rotti/collins:2020,clark/etal:2021,NPIPE:2022}, we model $C_\ell^{EB,\mathrm{dust}}$ by assuming that the dust $EB$ is proportional to the observed dust $TB$, that is, $C_\ell^{EB,\mathrm{dust}}/C_\ell^{EE,\mathrm{dust}}\propto C_\ell^{TB,\mathrm{dust}}/C_\ell^{TE,\mathrm{dust}}$. Specifically, we use
\begin{equation}
    \label{eq:dust_ansatz}
     C^{EB, \text{dust}}_\ell = A_\ell C^{EE, \text{dust}}_\ell \sin \left(4\psi_\ell\right)\,,
\end{equation}
where $A_\ell\ge 0$ is a free amplitude parameter and
\begin{equation}
     \psi_\ell = \frac{1}{2}\arctan\left(\frac{C^{TB,  \text{dust}}_\ell}{C^{TE, \text{dust}}_\ell  }\right)\,.
\end{equation}
We calculate $\psi_{\ell}$ from smoothing each of the symmetrized $TB$ and $TE$ spectra of the 353\,GHz A and B split maps using a one-dimensional Gaussian filter. The parameter $A_\ell$ is expected to depend on $\ell$ weakly. Following Refs.~\cite{NPIPE:2022,Eskilt:2022wav}, we sample $A_\ell$ for 4 ranges in $51 \leq \ell \leq 130$, $131 \leq \ell \leq 210$, $211 \leq \ell \leq 510$, and $511 \leq \ell \leq 1490$ with flat positive priors.

Unlike Refs.~\cite{NPIPE:2022,Eskilt:2022wav}, we do not relate Eq.~\eqref{eq:dust_ansatz} to the effective angle of the foreground $EB$, $\gamma_\ell$ (cf. Eq.~(2) of Ref.~\cite{NPIPE:2022}). We instead use Eq.~\eqref{eq:dust_ansatz} directly in our analysis. This procedure is more reliable because it does not require a small-angle approximation or multiplying a noisy factor ${C^{EE, \textrm{dust}}_\ell/(C^{EE, \textrm{dust}}_\ell-C^{BB, \textrm{dust}}_\ell)}$ taken from the 353\,GHz channels.

We define matrices
\begin{align}
    \mathbf{A}_{\ell, ij} = &\left[- \Lambda_{\ell}^T(\alpha_i,\alpha_j)\mathbf{\Lambda}_{\ell}^{-1}(\alpha_i, \alpha_j),\, 1\right],\\
    \nonumber
    \textbf{B}_{\ell, ij} =&\bigg[R^T(\alpha_i+\beta_i,\alpha_j+\beta_j)-\Lambda_\ell^T(\alpha_i,\alpha_j)\mathbf{\Lambda}_\ell^{-1}(\alpha_i, \alpha_j)\\
    &\cdot\textbf{R}(\alpha_i+\beta_i, \alpha_j+\beta_j)\bigg]\,,
\end{align}
where
\begin{align}
    \mathbf{\Lambda}_\ell(\alpha_i, \alpha_j) &= \mathbf{R}(\alpha_i, \alpha_j) + \mathbf{D}(\alpha_i, \alpha_j)\mathbf{F}_\ell,\\
    \Lambda_\ell^T(\alpha_i, \alpha_j) &= R^T(\alpha_i, \alpha_j) + D^T(\alpha_i, \alpha_j)\mathbf{F}_\ell,\\
    \mathbf{D}(\theta_i, \theta_j) &= \begin{bmatrix}
      -\cos(2\theta_i)\sin(2\theta_j) & -\sin(2\theta_i)\cos(2\theta_j) \\
        \phantom{-}\sin(2\theta_i)\cos(2\theta_j) & \phantom{-}\cos(2\theta_i)\sin(2\theta_j)
    \end{bmatrix},\\
    D(\theta_i, \theta_j) &= \begin{bmatrix}
        \phantom{-}\cos(2\theta_i)\cos(2\theta_j)\\
        -\sin(2\theta_i)\sin(2\theta_j)
    \end{bmatrix}\,,\\
    \mathbf{F}_\ell &= A_\ell \sin(4\psi_\ell)\begin{bmatrix}
        1 & 0\\
        1 & 0
    \end{bmatrix}\,.
\end{align}
Here, the matrix $\mathbf{F}$ is defined differently from that given in Refs. \cite{NPIPE:2022,Eskilt:2022wav} to simplify the expression.

We bin the observed power spectra with a bin size $\Delta \ell = 20$, and limit the range of multipoles to $\ell_{\textrm{min}}\le \ell\le \ell_{\textrm{max}}$ with $\ell_{\textrm{min}} = 51$ and $\ell_{\textrm{max}} = 1490$. We thus use $N_{\textrm{bin}}=72$ bins. The bin size and multipole range are the same as in the previous work~\cite{PlanckIntXLIX,minami/komatsu:2020b,NPIPE:2022,Eskilt:2022wav}. The results are robust against changes in $\ell_{\textrm{min}}$ and $\ell_{\textrm{max}}$. 

To find the posterior distributions of the sampled parameters ($\alpha_i$, $\beta_i$, $A_\ell$), we use a Metropolis Markov Chain Monte Carlo sampler to evaluate
\begin{equation}
    \ln L = -\frac{1}{2} \sum_{b=1}^{N_{\text{bin}}} \left(\vec{v}_b^T \textbf{M}_b^{-1}\vec{v}_b + \ln |\textbf{M}_b|\right)\,,
\end{equation}
where $b$ is the bin number, $\textbf{M}_b$ is the binned covariance matrix, and
$\vec{v}^T_b \equiv \textbf{A}\vec{C}^{\textrm{o}}_b - \textbf{B}\vec{C}^{\Lambda \text{CDM}}_b$. The unbinned covariance matrix is given by $\textbf{M}_\ell = \textbf{A}\text{Cov}(\vec{C}^{\textrm{o}}_\ell, \vec{C}^{\textrm{o}}_\ell)\textbf{A}^T$. We bin $\text{Cov}(\vec{C}^{\textrm{o}}_\ell, \vec{C}^{\textrm{o}}_\ell)$ as \cite{minami/komatsu:2020}
\begin{equation}
     \text{Cov}(C^{XY}_b, C^{ZW}_b) = \frac{1}{\Delta \ell ^ 2} \sum_{\ell \in b}  \text{Cov}(C^{XY}_\ell, C^{ZW}_\ell)\,,
\end{equation}
where we use an approximate covariance for each $\ell$,
\begin{equation}
    \label{eq:single-multipole-cov}
    \text{Cov}(C^{XY}_\ell, C^{ZW}_\ell) = \frac{C^{XZ,\textrm{o}}_\ell C^{YW,\textrm{o}}_\ell +  C^{XW,\textrm{o}}_\ell C^{YZ,\textrm{o}}_\ell}{(2\ell+1)f_{\textrm{sky}}}\,,
\end{equation}
with $f_{\textrm{sky}}$ being the fraction of sky used for the analysis.

We avoid using $EB$ terms in the right-hand side of Eq.~\eqref{eq:single-multipole-cov} due to the statistical fluctuations of $C_\ell^{EB,\textrm{o}}$. 
We thus set $\text{Cov}(C^{E_iB_j}_\ell, C^{E_pB_q}_\ell) = C^{E_iE_p}_\ell C^{B_jB_q}_\ell/[(2\ell+1)f_{\textrm{sky}}]$ where $i, j, p, q$ denote different maps. We also do not use off-diagonal elements in the covariance matrix of Eq.~\eqref{eq:single-multipole-cov}. 

We use \texttt{PolSpice}\footnote{\url{http://www2.iap.fr/users/hivon/software/PolSpice/}} \citep{Chon:2003gx} to get the observed power spectra in $\vec{C}^{\textrm{o}}_\ell$. 
We use 14 \planck\ and 10 \wmap\ maps, which give $24\cdot 23 = 552$ cross-power spectra, whereas auto-power spectra are excluded.

Our baseline result is based on the largest, nearly full-sky coverage mask used in Ref.~\cite{Eskilt:2022wav}. This mask excludes pixels in which the intensity of a carbon-monoxide (CO) line is stronger than  $45\,\textrm{K}_{\textrm{RJ}}\, \textrm{km} \,\textrm{s}^{-1}$. The CO emission is not polarized, but it could induce intensity-to-polarization leakage. Although the CO line exists only in some of the HFI maps, we apply the same CO mask to all the maps to simplify the analysis. The mask also excludes the locations of known polarized point sources. Specifically we use the union of the point-source masks of all the polarized \planck\ maps.

We calculate the sky fraction using \cite{Hivon:2002, Challinor:2004pr}
\begin{equation}
    f_{\textrm{sky}} = \frac{1}{N_{\text{pix}}}\frac{\left(\sum^{N_{\text{pix}}}_{i=1}w_i^2\right)^2}{ \sum^{N_{\text{pix}}}_{i=1}w_i^4},
\end{equation}
where $N_\text{pix}$ is the number of pixels and $w_i$ is the weight of the apodized mask at the $i$th pixel. We find $f_{\textrm{sky}} = 0.92$ for the baseline CO and point-source mask.

To explore the dependence of $\alpha_i$ and $C_\ell^{EB,\mathrm{dust}}$ on the mask, we also use a 30\% Galactic mask in union with the CO and point-source masks. The sky fraction is $f_\textrm{sky}=0.62$. These two masks correspond to the largest and smallest $f_{\textrm{sky}}$ used in Ref.~\cite{Eskilt:2022wav}.

The code to reproduce the results of this paper is publicly available\footnote{\url{https://github.com/LilleJohs/Cosmic_Birefringence}}.

\section{\label{sec:results}Results}
\begin{table}
\centering
\begin{tabular}{|cc |c | c  |  c}
\hline
\multicolumn{2}{|c|}{} & $f_{\textrm{sky}}=0.92$& $f_{\textrm{sky}}=0.62$\\ 
\hline
\multicolumn{4}{|c|}{\multirow{2}{*}{With the filamentary dust $EB$ model}}\\ \\
\hline
\planck\ HFI&$\alpha_i \neq 0$ &  $0.36^\circ \pm 0.11^\circ$&    $0.29^\circ \pm 0.28^\circ$     \\
\hline
\planck\ HFI+LFI& \multirow{2}{*}{$\alpha_i \neq 0$} &\multirow{2}{*}{$\mathbf{0.342^{\circ\,+0.094^\circ}_{\phantom{\circ\,}-0.091^\circ}}$} & \multirow{2}{*}{$0.37^\circ\pm0.14^\circ$} \\
+\wmap&&& \\
\hline
\multicolumn{4}{|c|}{\multirow{2}{*}{Without the filamentary dust $EB$ model}}\\ \\
\hline
\multirow{2}{*}{\planck\ HFI}& $\alpha_i=0$ & $0.288^\circ \pm 0.032^\circ$&   $\phantom{-}0.330^\circ \pm 0.035^\circ$ \\
&$\alpha_i\neq 0$ & $0.30^\circ \pm 0.11^\circ$&    $-0.25^\circ \pm 0.23^\circ$     \\
\hline
\planck\ HFI+LFI& $\alpha_i = 0$   & $0.298^\circ \pm 0.032^\circ$&   $0.343^\circ \pm 0.035^\circ$\\
+\wmap &  $\alpha_i \neq 0$ & $0.288^\circ \pm 0.091^\circ$ & $0.18^\circ\pm0.14^\circ$ \\
\hline
\end{tabular}
\caption{Measurements of the cosmic birefringence angle, $\beta$. The baseline result is shown in bold face. For ``$\alpha_i \neq 0$'' we jointly sample $\beta$, the miscalibration angles, $\alpha_i$, and the dust $EB$ amplitude, $A_\ell$ [Eq.~\eqref{eq:dust_ansatz}], whereas for ``$\alpha_i=0$'' we only sample $\beta$. The numbers in the third and seventh row (\planck\ HFI, $\alpha_i\ne 0$) are taken from Ref.~\cite{NPIPE:2022}.}
\label{table:measurements} 
\end{table}

First, we assume $\alpha_i=0$ and measure $\beta = 0.288^\circ \pm 0.032^\circ$ and $0.330^\circ\pm 0.035^\circ$ from the \planck\ HFI data for $f_\mathrm{sky}=0.92$ and 0.62, respectively (see the 6th row in Table~\ref{table:measurements}). We find similar results for other $f_\mathrm{sky}$. As the CMB can only determine the sum of $\beta$ and miscalibration angles, this result shows that the $EB$ signal is isotropic in the sky, and we robustly measure $\bar\alpha+\beta\simeq 0.3^\circ$ regardless of $f_\mathrm{sky}$, where $\bar\alpha$ is some suitable average value of $\alpha_i$ for the HFI. This result is not affected by $C_\ell^{EB,\mathrm{fg}}$ and is consistent with the \planck\ team's result performed on foreground-cleaned maps~\cite{PlanckIntXLIX}. The foreground emission is not responsible for this signal.

Using all the \wmap\ and \planck\ polarization data and still assuming $\alpha_i=0$, we measure $\beta = 0.298^\circ \pm 0.032^\circ$ and $0.343^\circ\pm 0.035^\circ$ for $f_\mathrm{sky}=0.92$ and 0.62, respectively (second-last row). The statistical power of the data is sufficient to make a significant detection of $\beta$, provided that we know $\alpha_i$. 

As both the \wmap\ and \planck\ have flown and ended the missions already, we cannot precisely measure the miscalibration angles of their detectors anymore. We thus rely on the foreground to measure $\alpha_i$ as done in Refs.~\cite{minami/komatsu:2020b,NPIPE:2022,Eskilt:2022wav}. When $C_\ell^{EB,\mathrm{fg}}$ is ignored in the analysis, the \wmap\ and \planck\ data yield $\beta = 0.288^\circ \pm 0.091^\circ$ for $f_\mathrm{sky}=0.92$. This is more precise than the \planck\ HFI result, $\beta=0.30^\circ\pm 0.11^\circ$~\cite{NPIPE:2022}, showing the additional information gained from the LFI and \wmap\ data. 

Still ignoring $C_\ell^{EB,\mathrm{fg}}$, we find that the Galactic mask reduces the value of $\beta$ to $0.18^\circ \pm 0.14^\circ$ for $f_\mathrm{sky}=0.62$, but not as much as to $-0.25^\circ\pm 0.23^\circ$ found for the \planck\ HFI-only result~\cite{NPIPE:2022}. The inclusion of the cross-power spectra with low frequency bands, in which the intensity of polarized dust emission is much weaker, reduces the impact of $C_\ell^{EB,\mathrm{dust}}$ and significantly increases the measured value of $\beta$. The \planck\ LFI+HFI analysis reported in Ref.~\cite{Eskilt:2022wav} gave $\beta = 0.14^\circ \pm 0.17^\circ$ for $f_{\textrm{sky}}=0.62$. The inclusion of the \wmap\ channels further increases the mean value and tightens the uncertainty of $\beta$.

\begin{figure}
\centering
\includegraphics[width=\linewidth]{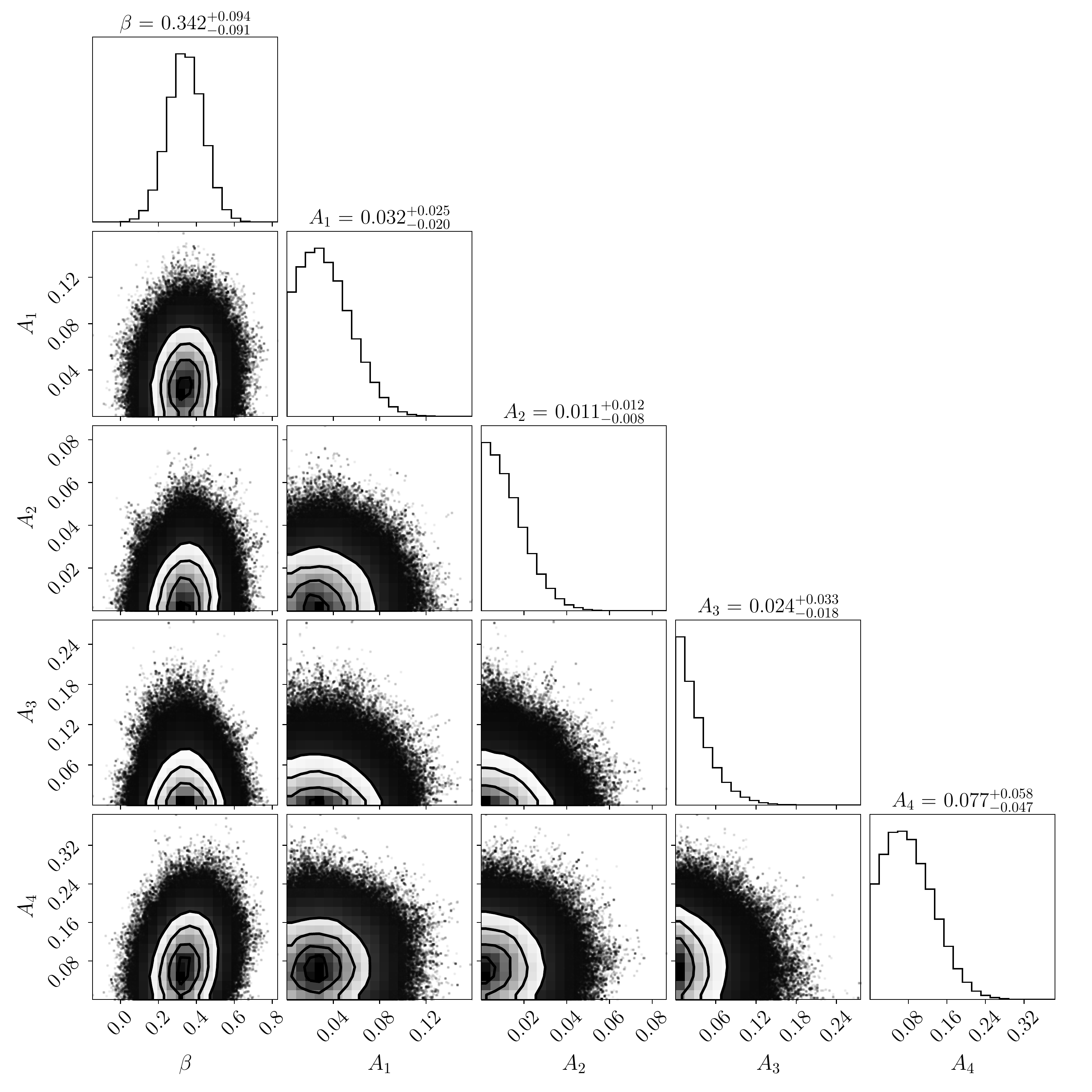}
\caption{\label{fig:beta_full} Posterior distributions of the cosmic birefringence angle, $\beta$, and the dust $EB$ amplitude, $A_\ell$ [Eq.~\eqref{eq:dust_ansatz}], in 4 bins for nearly full-sky data ($f_\mathrm{sky}=0.92$; the 4th row in Table~\ref{table:measurements}). The miscalibration angles, $\alpha_i$, are jointly sampled with $\beta$ and $A_\ell$ but not shown here. See Fig.~\ref{fig:alpha_full} for the 1-d marginalized posterior distribution of $\alpha_i$.
}
\end{figure}
\begin{figure}
\centering
\includegraphics[width=\linewidth]{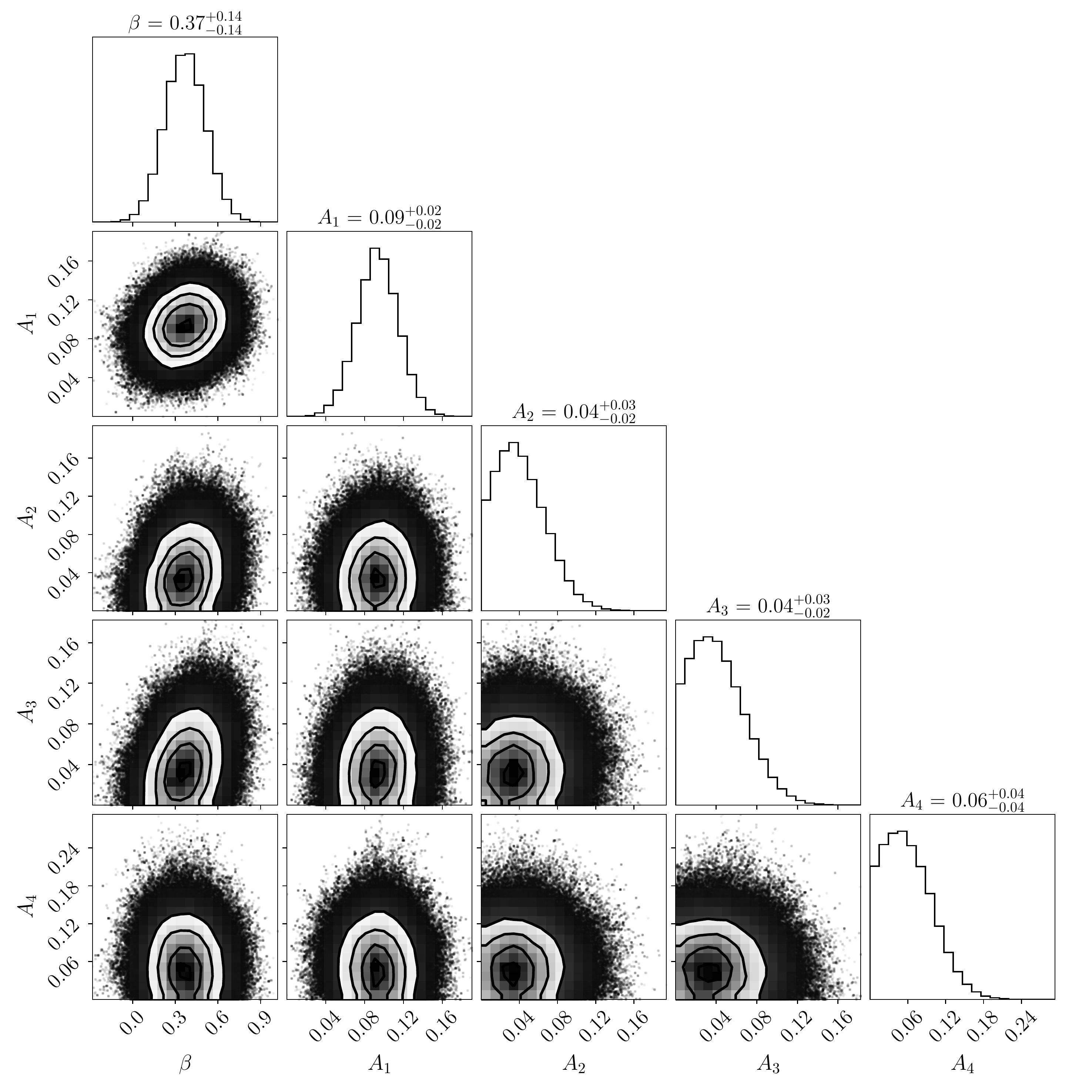}
\caption{\label{fig:beta_part}Same as Fig.~\ref{fig:beta_full} but for $f_\mathrm{sky}=0.62$.
}
\end{figure}

We now account for $C_\ell^{EB,\mathrm{fg}}$. 
The foreground emission in the \planck\ LFI bands and similar frequency bands of \wmap\ is dominated by synchrotron rather than by dust. Unlike for dust, there is no evidence for the intrinsic $EB$ correlation of synchrotron emission~\cite{Martire:2021gbc}. We thus ignore $C_\ell^{EB,\mathrm{synch}}$, but use $C_\ell^{EB,\mathrm{dust}}$ presented in Eq. \eqref{eq:dust_ansatz} to model the dust $EB$ correlation and apply it only to the HFI maps of \planck\ and the 94~GHz maps of \wmap.

\begin{figure}
\centering
\includegraphics[width=\linewidth]{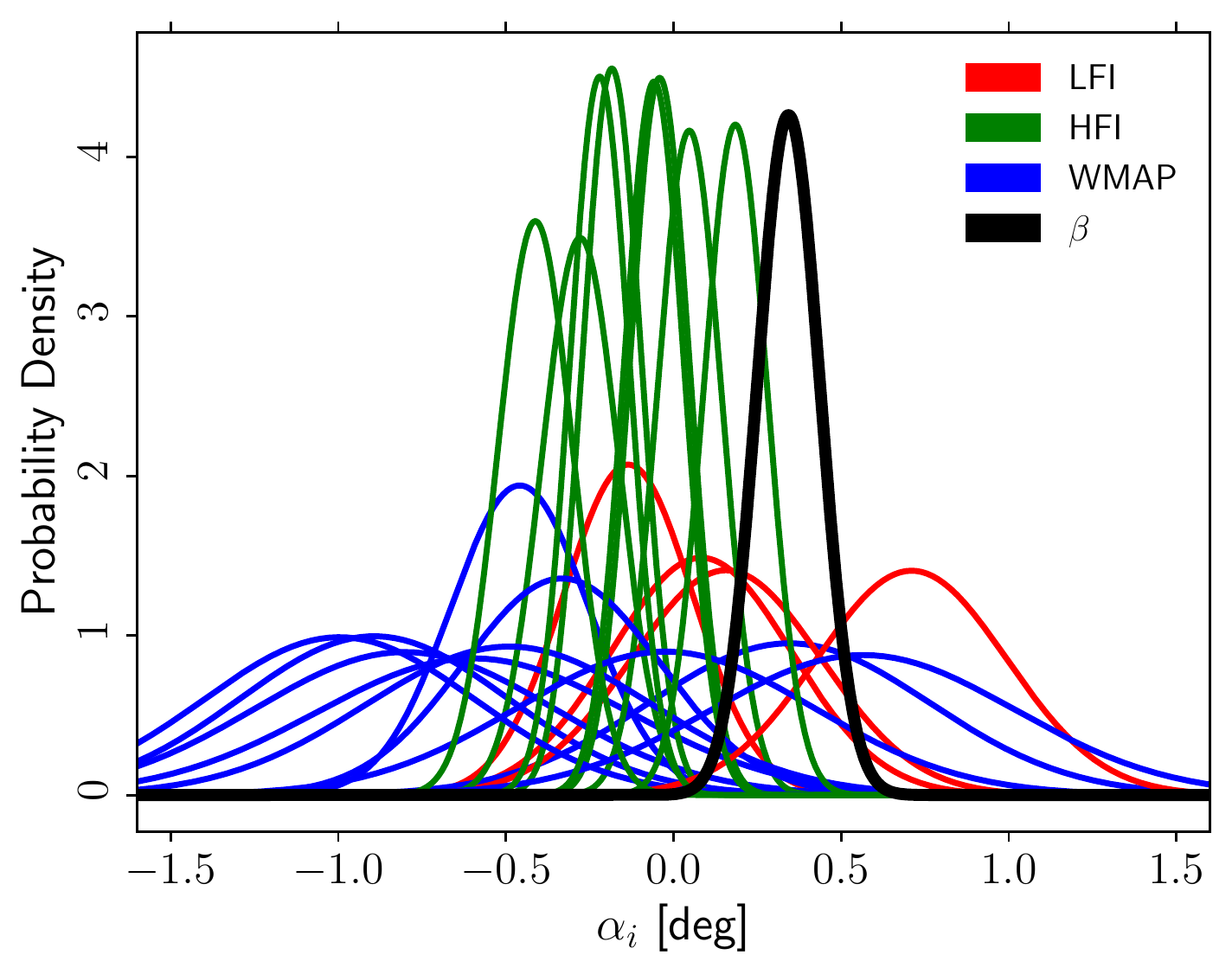}
\caption{\label{fig:alpha_full}Marginalized posterior distributions of 22 miscalibration angles, $\alpha_i$, for the baseline result. The distribution of $\beta$ (thick black line) is the same as in the top-left corner of Fig.~\ref{fig:beta_full}.
}
\end{figure}

Jointly sampling a frequency-independent $\beta$, 22 miscalibration angles $\alpha_i$, and the dust $EB$ amplitudes $A_\ell$ in 4 ranges of $\ell$, we measure $\beta = 0.342^{\circ\,+0.094^\circ}_{\phantom{\circ\,}-0.091^\circ}$ and $0.37^\circ\pm 0.14^\circ$ for $f_\textrm{sky}=0.92$ and 0.62, respectively (the 4th row in Table~\ref{table:measurements}). The former is our baseline result, which excludes $\beta=0$ at 99.987\%~C.L. The latter agrees with the former within $1\sigma$ and excludes $\beta=0$ at 99.5\%~C.L. 

In Figs.~\ref{fig:beta_full} and \ref{fig:beta_part} we show the posterior distributions of $\beta$ and $A_\ell$ in 4 bins for $f_\textrm{sky}=0.92$ and 0.62, respectively. The dust $EB$ amplitudes are consistent with zero except for the first multipole bin ($51\le\ell\le 130$) for $f_\mathrm{sky}=0.62$.

In Fig.~\ref{fig:alpha_full} we show the 1-dimensional marginalized posterior distributions of 22 miscalibration angles, $\alpha_i$, for the baseline result. The distributions for the \wmap\ and \planck\ LFI are broader than those of the \planck\ HFI, as expected from the signal-to-noise ratio. The measured $\alpha_i$ are in agreement with the reported calibration uncertainties, $1.5^\circ$ for \wmap\ \cite{WMAP:2011} and $1^\circ$ for \planck\ \cite{Rosset2010}.

We now introduce a frequency-dependent cosmic birefringence signal as $\beta(\nu)=\beta_0(\nu/150~\mathrm{GHz})^n$~\cite{Eskilt:2022wav}. Using the \wmap\ and \planck\ data and accounting for $C_\ell^{EB,\mathrm{dust}}$, we find $\beta_0 = 0.30^\circ \pm 0.10^\circ$ and $n = -0.20^{+0.41}_{-0.39}$. 
Our measurements are thus consistent with a frequency-independent cosmic birefringence signal predicted by the axionlike field. 

\begin{figure}
\centering
\includegraphics[width=\linewidth]{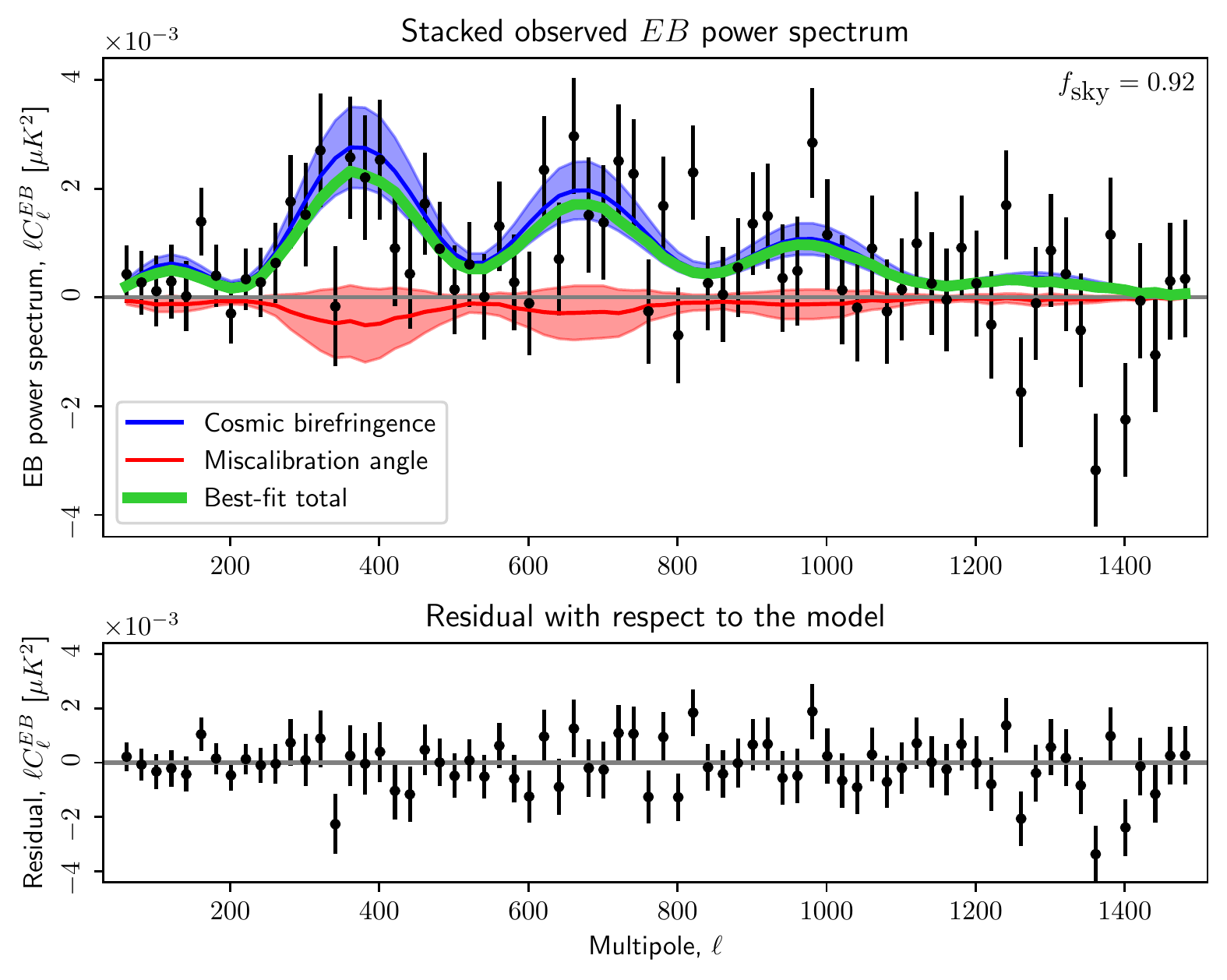}
\caption{\label{fig:stacked-eb} Stacked observed $EB$ power spectrum (upper) and residuals with respect to the best-fitting model (lower) for nearly full-sky data ($f_{\textrm{sky}}=0.92$). The beam transfer and pixel window functions are not deconvolved. The blue and red shaded areas show the $1\sigma$ bands for the cosmic birefringence and miscalibration contributions, respectively. The green solid line shows the total best-fitting model.
}
\end{figure}

These results support the following hypothesis. Synchrotron emission has little to no intrinsic $EB$ correlations that could bias the measurement of $\beta$, whereas dust contains $EB$ correlations which bias the measurement when not accounted for, especially for smaller $f_\mathrm{sky}$. 
Although no synchrotron $EB$ has been found so far \cite{Martire:2021gbc}, more evidence is needed to rule out synchrotron $EB$ as one of the culprits. The C-BASS and QUIJOTE experiments will give us a better understanding of synchrotron in the near future \cite{dickinson/etal:2018, QUIJOTE:2018ntj}.

\begin{figure}
\centering
\includegraphics[width=\linewidth]{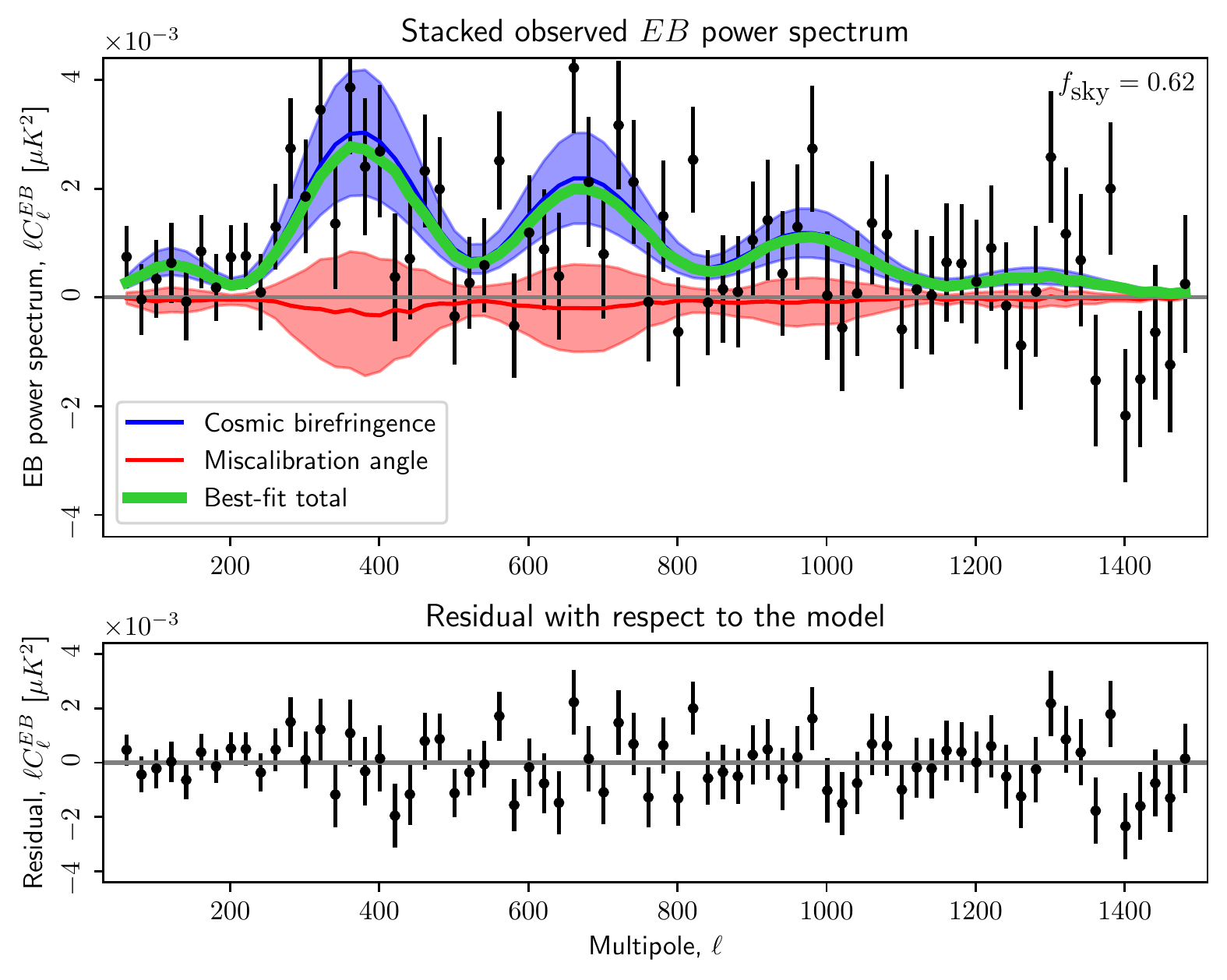}
\caption{\label{fig:stacked-eb-062} Same as Fig.~\ref{fig:stacked-eb} but for $f_{\textrm{sky}}=0.62$.
}
\end{figure}

Finally, we show the average of the observed $EB$ power spectra (``stacked observed $EB$ power spectrum'') with the uncertainty in Fig.~\ref{fig:stacked-eb} and Fig.~\ref{fig:stacked-eb-062} for $f_\textrm{sky}=0.92$ and 0.62, respectively.  We calculate the inverse-variance weighted average of the observed $EB$ power spectra from
\begin{equation}
    \textrm{E}\left(C^{EB, \textrm{o}}_b \right) = \frac{\vec{1} \cdot \textbf{M}^{-1}_b \cdot \vec{v}_b}{\vec{1} \cdot \textbf{M}^{-1}_b \cdot \vec{1}}\,,
\end{equation}
and the variance from
\begin{equation}
    \textrm{Var}\left(C^{EB, \textrm{o}}_b \right) = \frac{1}{\vec{1} \cdot \textbf{M}^{-1}_b \cdot \vec{1}}\,.
\end{equation}
We use $\alpha_i=0$ and $\beta = 0$ in $\vec{v}_b$ to get the stacked $EB$ power spectrum shown in the black points with error bars in the upper panels of Fig.~\ref{fig:stacked-eb} and Fig.~\ref{fig:stacked-eb-062}. We find similar stacked $EB$ power spectra for both sky fractions, which are expected because 
similar values of $\beta$ are found when assuming $\alpha_i=0$ (see the 4th row of Table~\ref{table:measurements}).

The blue $1\sigma$ bands show the $EB$ power spectra as predicted by the baseline cosmic birefringence angle, $\beta = 0.342^{\circ\,+0.094^\circ}_{\phantom{\circ\,}-0.091^\circ}$ for $f_{\textrm{sky}}=0.92$, and $\beta = 0.37^\circ \pm 0.14^\circ$  for $f_{\textrm{sky}}=0.62$. We fix $\alpha_i=0$ in the covariance matrix, $\textbf{M}_b$, for this case. The red $1\sigma$ bands show the contributions from $\alpha_i$ where $\alpha_i$ is included in $\textbf{M}_b$. The smaller sky fraction increases the uncertainty on the contribution from $\alpha_i$, but still yields consistent results.

In the lower panels we show the residual with respect to the best-fitting $A_\ell$, $\alpha_i$ and $\beta$ model. The $\chi^2$ for the degree of freedom of 72 is 65.3 and 65.8 for $f_\textrm{sky}=0.92$ and 0.62, respectively. We thus conclude that the residuals are consistent with null.

\section{\label{sec:conclusion}Conclusions}
\begin{table}
\centering
\begin{tabular}{|c |c | c|}
\hline Datasets &  $\beta$ & Dust $EB$ model\\ 
\hline
    \planck\ PR3 HFI \citep{minami/komatsu:2020b} & $0.34^\circ \pm 0.14^\circ$ & No \\
     \planck\ PR4 HFI \citep{NPIPE:2022}& $0.30^\circ \pm 0.11^\circ$ & No  \\
    \planck\ PR4 HFI \citep{NPIPE:2022}& $0.36^\circ \pm 0.11^\circ$& Yes  \\
     \planck\ PR4 HFI + LFI \citep{Eskilt:2022wav}& $0.33^\circ \pm 0.10^\circ$& No  \\
     \planck\ PR4 + \wmap\ & $0.342^{\circ\,+0.094^\circ}_{\phantom{\circ\,}-0.091^\circ}$& Yes  \\
\hline
\end{tabular}
\caption{Previous measurements of cosmic birefringence at nearly full-sky that adopts the method of Refs.~\citep{minami/etal:2019, minami/komatsu:2020}. The right column indicates if the filamentary dust model of $C^{EB, \textrm{dust}}_\ell$ [Eq.~\eqref{eq:dust_ansatz}] was applied to the high-frequency maps where dust is the dominating foreground contribution. The last row displays this work.}
\label{table:previous-measurements} 
\end{table}
We have presented new constraints on the cosmic birefringence angle, $\beta$, from a joint analysis of the $EB$ power spectra of the \wmap\ 9-year maps~\cite{WMAP:2013a} and the \planck\ PR4 LFI and HFI maps~\cite{PlanckIntLVII}, which cover a wide range of frequencies from $\nu=23$ to 353~GHz.
We used the method based on Refs.~\cite{huffenberger/rotti/collins:2020,clark/etal:2021,NPIPE:2022} to account for the potential impact of the intrinsic $EB$ correlation of polarized dust emission on the determination of instrumental miscalibration angles, $\alpha_i$, for $\nu\ge 94$~GHz. Marginalizing over $\alpha_i$ and the dust $EB$ amplitudes, we measure $\beta = 0.342^{\circ\,+0.094^\circ}_{\phantom{\circ\,}-0.091^\circ}$ (68\%~C.L.) for nearly full-sky data, excluding zero at 99.987\%~C.L. This is consistent with, and more precise than, the previous results from the \planck\ data~\cite{minami/komatsu:2020b,NPIPE:2022,Eskilt:2022wav}, and corresponds to the statistical significance of $3.6\sigma$.

The consistent results of the joint analysis reinforce the conclusion of the \planck\ HFI analysis~\cite{NPIPE:2022,diego-palazuelos/etal:2022} that  errors due to instrumental systematics are smaller than statistical errors. In Table~\ref{table:previous-measurements} we show previous measurements of cosmic birefringence where there is a clear trend of increased statistical significance when new datasets are added or a filamentary dust model for $EB$ is included. 

If we remove the Galactic plane from the analysis, we find $\beta=0.37^\circ\pm 0.14^\circ$ for $f_\mathrm{sky}=0.62$, excluding zero at 99.5\%~C.L. We thus find consistent signals of cosmic birefringence for both sky fractions. We also find that adding the cross-power spectra with lower frequency data from \wmap\ and \planck\ LFI reduces the impact of polarized dust emission on $\alpha_i$ compared to the HFI-only analysis~\cite{Eskilt:2022wav}. We have not accounted for the intrinsic $EB$ correlation of synchrotron emission which dominates at low frequencies because there is no evidence for it~\cite{Martire:2021gbc, Martire:2022qkr}.

We find no evidence for frequency dependence of $\beta$. For $\beta(\nu)\propto \nu^n$, we measure $n = -0.20^{+0.41}_{-0.39}$ when accounting for $C^{EB, \textrm{dust}}_\ell$. This is consistent with $n=0$ predicted by the axionlike field [Eq.~\eqref{eq:lagrangian}] but disfavors, for example, $n=-2$ predicted by the Faraday rotation effect from the intergalactic (including primordial) or interstellar magnetic field.

A better understanding of the foreground emission is needed to completely rule out the foreground $EB$ as the culprit. However, the best way forward is to improve upon the calibration work rather than relying on the foreground emission to determine $\alpha_i$. If the calibration accuracy reaches $\pm 0.06^\circ$, $\beta$ can be reliably detected with a statistical significance of $>5\sigma$. We also need confirmation from more independent datasets to completely rule out the (unknown) systematics of \wmap\ and \planck. 

To this end, both on-going and future ground-based~\cite{adachi/etal:2020b,choi/etal:2020,dutcher/:2021,BICEP:2021,dahal/etal:2022,QUBIC:2022,SimonsObservatory:2019,SPO:2020,CMB-S4:2019}, balloon-borne~\cite{SPIDER:2022,LSPE:2021}, and space-borne~\cite{LiteBIRD:2022,NASAPICO:2019} experiments are expected to lead to a convincing discovery (or otherwise) of cosmic birefringence.
If proven to be a cosmological signal, isotropic cosmic birefringence would have a profound impact on cosmology, particle physics, and quantum gravity~\cite{fujita/etal:2021a,fujita/etal:2021b,takahashi/yin:2021,mehta/etal:2021,nakagawa/takahashi/yamada:2021,alvey/escudero:2021,choi/etal:2021,obata:prep,Gasparotto:2022uqo,Nakatsuka:2022epj,kitajima/etal:prep}.

\begin{acknowledgments}
We thank Patricia Diego-Palazuelos, Hans Kristian Eriksen, Kris M. G\'orski, Yuto Minami, Matthieu Tristram, Duncan Watts, and Ingunn Wehus for useful discussions and comments on the paper.
This work was supported in part by the European Research Council (ERC) under the Horizon 2020 Research and Innovation Programme (Grant agreement No.~819478), JSPS KAKENHI Grants No.~JP20H05850 and No.~JP20H05859, and the Deutsche Forschungsgemeinschaft (DFG, German Research Foundation) under Germany's Excellence Strategy - EXC-2094 - 390783311. This work has also received funding from the European Union's Horizon 2020 research and innovation programme under the Marie Sk\l odowska-Curie grant agreement No.~101007633.
The Kavli IPMU is supported by World Premier International Research Center Initiative (WPI), MEXT, Japan. We acknowledge the use of the Legacy Archive for Microwave Background Data Analysis (LAMBDA), part of the High Energy Astrophysics Science Archive Center (HEASARC). HEASARC/LAMBDA is a service of the Astrophysics Science Division at the NASA Goddard Space Flight Center. \planck\ is a project of the European Space Agency (ESA) with instruments provided by two scientific consortia funded by ESA member states and led by Principal Investigators from France and Italy, telescope reflectors provided through a collaboration between ESA and a scientific consortium led and funded by Denmark, and additional contributions from NASA (USA). Some of the results in this paper have been derived using the \texttt{HEALPix} package~\cite{gorski/etal:2005}.
\end{acknowledgments}
\bibliography{references}
\end{document}